%Paper: 9109020
%From: leclair@strange.tn.cornell.edu (Andre LeClair)
%Date: Wed, 11 Sep 91 14:57:14 EDT

\input harvmac.tex
%%%%%%%%%%%%%%%%%%%%%%%%%%%%%%%%%%%%%%%%%%%%%%%%%%%%%%%%%%%%%%%%%
%%%%%  Infinite Quantum Group Symmetry in 2d Quantum Field Theory
%%%%%     by Andre LeClair
%%%%%%%%%%%%%%%%%%%%%%%%%%%%%%%%
\noblackbox

%%%%%%%%%%%%%%%%%%%%%%%%%%%%%%%%%%%%%%%%%%%%%%%%%%%%%%%%%%%%%%%
%
%		DEFINITIONS FOR TEX
%
%%%%%%%%%%%%%%%%%%%%%%%%%%%%%%%%%%%%%%%%%%%%%%%%%%%%%%%%%%%%%%%
%

%
%\def\e{\'e}
%\def\ee{\`e}
%%%%%%%%%%%%%%%%%%%DEFINITIONS%%%%%%%%%%%%%%%%%%%%%%%%%%%%%%%%%
%

\def\bra#1{{\langle #1 |  }}
\def\lb{ \left[ }
\def\rb{ \right]  }

\def\bar{\overline}
\def\hat{\widehat}
\def\*{\star}
\def\[{\left[}
\def\]{\right]}
\def\({\left(}		\def\BL{\Bigr(}
\def\){\right)}		\def\BR{\Bigr)}
	\def\BBL{\lb}
	\def\BBR{\rb}
%
%%%%%%%%%%%%%%%%%%%%%%%%%%%%%%%%%%%%%%%%%%%%%%%%%%%%%%%%%%%%%%%
%
\def\zb{{\bar{z} }}
\def\frac#1#2{{#1 \over #2}}
\def\inv#1{{1 \over #1}}
\def\half{{1 \over 2}}
\def\d{\partial}

\def\ket#1{ | #1 \rangle}

\def\2pi{\hbox{$2\pi i$}}

\def\dsl{\raise.15ex\hbox{/}\kern-.57em\partial}
\def\Dsl{\,\raise.15ex\hbox{/}\mkern-.13.5mu D}
%
%%%%%%%%%%%%%%%%%%%%GREEK LETTERS%%%%%%%%%%%%%%%%%%%%%%%%%%%%%%
%
\def\th{\theta}		\def\Th{\Theta}
\def\ga{\gamma}		
\def\be{\beta}
\def\al{\alpha}
\def\ep{\epsilon}
\def\la{\lambda}	\def\La{\Lambda}
\def\de{\delta}		\def\De{\Delta}

\def\vphi{\varphi}
%
%%%%%%%%%%%%%%%%%%%CALIGRAPHIC LETTERS%%%%%%%%%%%%%%%%%%%%%%%%%
%
\def\CA{{\cal A}}

\def\CJ{{\cal J}}		
		
		\def\CR{{\cal R}}
		\def\CU{{\cal U}}

\def\2pi{\hbox{$2\pi i$}}

\def\dsl{\raise.15ex\hbox{/}\kern-.57em\partial}
\def\Dsl{\,\raise.15ex\hbox{/}\mkern-.13.5mu D}
%
%%%%%%%%%%%%%%%%%%%%GREEK LETTERS%%%%%%%%%%%%%%%%%%%%%%%%%%%%%%
%
%%%%%%%%%%%%%%% MATH CHARACTERS %%%%%%%%%%%%%%%%%%%%%%%%%%%%
%
\font\numbers=cmss12
%\font\numbers=cmu10 scaled\magstep1
\font\upright=cmu10 scaled\magstep1
\def\stroke{\vrule height8pt width0.4pt depth-0.1pt}
\def\topfleck{\vrule height8pt width0.5pt depth-5.9pt}
\def\botfleck{\vrule height2pt width0.5pt depth0.1pt}
\def\Zmath{\vcenter{\hbox{\numbers\rlap{\rlap{Z}\kern 0.8pt\topfleck}\kern
2.2pt
                   \rlap Z\kern 6pt\botfleck\kern 1pt}}}
\def\Qmath{\vcenter{\hbox{\upright\rlap{\rlap{Q}\kern
                   3.8pt\stroke}\phantom{Q}}}}
\def\Nmath{\vcenter{\hbox{\upright\rlap{I}\kern 1.7pt N}}}
\def\Cmath{\vcenter{\hbox{\upright\rlap{\rlap{C}\kern
                   3.8pt\stroke}\phantom{C}}}}
\def\Rmath{\vcenter{\hbox{\upright\rlap{I}\kern 1.7pt R}}}
\def\Z{\ifmmode\Zmath\else$\Zmath$\fi}
\def\Q{\ifmmode\Qmath\else$\Qmath$\fi}
\def\N{\ifmmode\Nmath\else$\Nmath$\fi}
\def\C{\ifmmode\Cmath\else$\Cmath$\fi}
\def\R{\ifmmode\Rmath\else$\Rmath$\fi}
%%%%%%%%%%%%%%%%%%%%%%%%%%%%%%%%%%%%%%%%%%%%%%%%%%%%%%%%%%%%%%%%%
 %%%%%%%%%%%%%%%%%% END OF DEFINITIONS %%%%%%%%%%%%%%%%%%%%%%
 %%%%%%%%%%%%%%%%%%%%%%%%%%%%%%%%%%%%%%%%%%%%%%%%%

%%%%%%%%%%%%%%%%%%%%%%%%%%%%%%%%%%%%%%%%%%%%%%%%%%%%%%%%%%%%%%%%%
%%%%%%%%%%%%%%%%%%REFERENCES%%%%%%%%%%%%%%%%%%%%%%%%%%%%%
%%%%%%%%%%%%%%%%%%%%%%%%%%%%%%%%%%%%%%%%%%%%%%%%%%%%%%%%%%

\def\Mat{S. D. Mathur, `Quantum Kac-Moody Symmetry in Integrable
Field Theories', preprint HUTMP-90-B/299 (1990).}

\def\form{F. A. Smirnov, {\it Form Factors in Completely Integrable
Models of Quantum Field Theory}, World Scientific\semi
A. N. Kirillov and F. A. Smirnov, Phys. Lett. 198B (1987) 506;
Zap. Nauch. Semin. LOMI 164 (1987) 80.}

\def\Mand{S. Mandelstam, Phys. Rev. D11 (1975) 3026.}

%
%
%
%%%%%%%%%%%%%%%%%%%%%%%%%%%%%%%%%%%%%%%%%%%%%%%%%%%%%%%%%%%%%%%%%%%%%%

%

\def\Luschi{M. L\"uscher, Nucl. Phys. B135 (1978) 1.}

\def\KaTh{M. Karowski and H. J. Thun, Nucl. Phys. B190 (1981) 61.}

\def\Yang{D. Bernard,
Commun. Math. Phys. 137 (1991) 191. }

\def\WOlive{E. Witten and D. Olive, Phys. Lett. B78 (1978) 97.}

%%%%%%%%%%%%%%%%%%%%%%%%%%%%%%%%%%%%%%%%%%%%%%%%%%%%%%%%%%%%%%%%%%%%%%%%%

%%%%%%%%%%%%%%%%%%%%%%%%%%%%%%%%%%%%%%%%%%%%%%%%%%%%%%%%%%%%%%%%%%%%%%%%%

%

%
%

%
%

%

%
%

%

%

%
%

%

%

%

%
%
%
%

%

%
%

%
%

%
%

%
%

%
%

%
%

%
%

%
%

%
%

%
%

%
%

%
%

%
%

%
%

%
%

%
%
%

%
%
%

%
%

%
%

%
%

%
%

%
%

%

%
%

%

\def\Drinfeld{V. G. Drinfel'd, Sov. Math. Dokl. 32 (1985) 254;
Sov. Math. Dokl. 36 (1988) 212. }
\def\Jimbo{M. Jimbo, Lett. Math. Phys. 10 (1985) 63; Lett. Math. Phys. 11
(1986) 247; Commun. Math. Phys. 102 (1986) 537.}

\def\Zamolodchikovi{A. B. Zamolodchikov and Al. B. Zamolodchikov, Annals
Phys. 120 (1979) 253.}
\def\Colemani{S. Coleman, Phys. Rev. D 11 (1975) 2088.}

\def\GoS{C. Gomez and G. Sierra, Phys. Lett. B240 (1990) 149;
Nucl. Phys. B352 (1991) 791.}
\def\Zamoiii{A. B. Zamolodchikov, Int. Journ. of Mod. Phys. A4 (1989) 4235;
in Adv. Studies in Pure Math. vol. 19 (1989) 641.}
%

%

%

%

%

%

%%%%%%%%%%%%%%%%%%%%%%%%%%%%%%%%%%%%%%%%%%%%%%%%%%%%%%%%%%%%%%%%%%%%%%%
%%%%%%%%%%%%%%%%%%%%%% END REFERENCES %%%%%%%%%%%%%%%%%%%%%%%%%%%%%%%%%%
%%%%%%%%%%%%%%%%%%%%%%%%%%%%%%%%%%%%%%%%%%%%%%%%%%%%%%%%%%%%%%%%%%%%%%
\Title{CLNS 91/1107}
{\vbox{\centerline{Infinite Quantum Group Symmetry   }
\centerline{ in 2d Quantum Field Theory} }}

\bigskip
\bigskip

\centerline{ANDR\' E LECLAIR}
\medskip\centerline{Newman Laboratory}
\centerline{Cornell University}
\centerline{Ithaca, NY  14853}

\vskip .3in

I describe how integrable quantum field theories in 2
spacetime dimensions are characterized by infinite
dimensional quantum group symmetries, namely the
q-deformations of affine Lie algebras, and their
Yangian limit.  These symmetries can provide a new
non-perturbative formulation of the theories.

\bigskip
\bigskip
\bigskip
\noindent
{\it Invited talk at the  XXth International Conference
on Differential Geometric Methods in Theoretical Physics, New York City,
June 1991.}
\Date{9/91}

%

%\Date{9/91}
%\draft
%
\def\hw{\Lambda}
\def\f#1#2#3{{ f^{#1#2#3} }}
\def\rhw{\rho_{\hw} }
\def\rhwo#1{{\rho_{\hw_#1} }}
\def\dis{\displaystyle}
%
%
%
%
%sample reference
%
%on an operator formulation of the superstring\ref\a{\sdual} .
%
%sample equations

%\eqn\one{
%\V 123 A\rangle_1 \vb_2 \vc_3 =
%\langle h_1 \left[ V_A (0)\right] h_2 \left[ V_B (0) \right]
%h_3 \left[ V_C (0) \right] \rangle . }
%
%
%
%\eqnn\two
%$$\eqalignno{
% h_1 (z ) &= z \cr
 %h_2 (z ) &= { 1\over {1-z } }&\two\cr
% h_3 (z ) &= { {z -1}\over {z} } .\cr}$$
%
%eq a,b,c etc
%\eqna\three
%$$\eqalignno{
%.............&\three {a} \cr
%..........&\three {b} \cr
%}$$

\def\formn{F. A. Smirnov, {\it Form Factors in Completely Integrable
Models of Quantum Field Theory}, World Scientific\semi
A. N. Kirillov and F. A. Smirnov, Phys. Lett. 198B (1987) 506;
Zap. Nauch. Semin. LOMI 164 (1987) 80.}

\def\ot{\otimes}

\def\m#1#2#3{m_{#1#2}^{#3} }
\def\u#1#2#3{\mu^{#1#2}_{#3} }
\def\ad#1{ {\rm ad}_{\displaystyle {#1}} }

\def\qo#1{Q_0^{#1}}
\def\qp#1{{Q_1^{#1}}}

\def\jua{{ J^a_\mu }}

\def\TH{\Theta}

\def\Th{\TH}

\def\Psib{{ \bar{\Psi} }}
\def\ad#1{{ {\rm ad}_{#1} }}
\def\vphi{\varphi}

\def\slqh{{ $\CU_q \( \hat{sl(2)}  \)$ }}

\def\phb#1#2{{ \phib^{(#1,#2)} }}
\def\Qb{{ \bar{Q} }}
\def\bQ{{ \bar{Q} }}
\def\Jb{{ \bar{J} }}
\def\a{{\rm ad}}

\def\vphib{{\bar{\varphi} }}

\def\R{{ \check{R} }}
\def\jua{ J^a_{\mu} }

\def\hb{ {\hat \beta} }

\def\slq{ {\hat {sl_q(2)} }}
\def\Psib{{\bar \Psi}}

\newsec{{\bf Introduction} }

In this talk I will review work done in a series of
papers with D. Bernard, G. Felder, and F. Smirnov
\ref\nlc{D. Bernard and A. LeClair, `Quantum Group
Symmetries and Non-Local Currents in 2D QFT', to
appear in Commun. Math. Phys.; Phys. Lett. B247 (1990) 309.}
\ref\form{A. LeClair and F. Smirnov, `Infinite Quantum
Group Symmetry of Fields in Massive 2D Quantum Field
Theory, to appear in Int. Journ. Mod. Phys.}
\ref\gio{G. Felder and A. LeClair, `Restricted
Quantum Affine Symmetry of Perturbed Minimal Conformal
Models', to appear in Proceedings of RIMS Research
Project 91, Int. Journ. Mod. Phys.}.
The subject concerns new symmetry structures of
integrable quantum field theories in 2 space-time
dimensions.

A class of quantum field theories of special interest
are those which possess asymptotic massive particle states.
The quantum mechanical scattering of these states is
described by the so-called S-matrix.   The importance of symmetry in
quantum field theory has long been recognized.
In 4 space-time dimensions, the possible symmetries of the
S-matrix are severely limited by the Coleman-Mandula
theorem, which states that such symmetries are necessarily
the tensor product of Poincar\'e symmetry with some internal
symmetry.  Though an important result, this theorem was
rather disappointing, since these allowed symmetries are
normally not large enough to provide a non-perturbative solution.
One of the hidden assumptions of the Coleman-Mandula theorem
is that the symmetry acts on multiparticle states as if
they were tensor products of 1-particle states.
More precisely if $Q$ is a conserved charge, then it was
assumed that its representation on a multiparticle state,
which is here denoted as $\De (Q)$, is given by
\eqn\Ii{
\Delta (Q) = Q\otimes 1\otimes \cdot\cdot \otimes 1 ~+~
1\otimes Q \otimes 1\otimes\cdot\cdot\otimes 1 ~+~ \cdot\cdot\cdot
{}~ + 1\otimes \cdot\cdot \otimes Q . }
In modern mathematical terminology, it was assumed that the
comultiplication was trivial.

\def\slq{{ \hat{sl_q (2)} }}

In 4 dimensions the only known exception to the Coleman-Mandula
theorem is supersymmetry, and it is due to the possibility of
fermionic currents in 4d.  However in lower dimensions, the
possibility of more exotic statistics of fields implies
there can be more non-trivial exceptions to the Coleman-Mandula
theorem.  The symmetries I will discuss in the following
are precisely of this nature.
Let me summarize some of the main features.  For simplicity
of the presentation I will limit myself to the sine-Gordon (SG)
and its Gross-Neveu Limit.  One can construct explicitly
some non-local conserved currents in the SG theory that generate
a q-deformation of sl(2) affine Lie algebra ($\slq$) \nlc .
The currents which generate this symmetry have non-trivial, though
abelian, braiding relations.  At a special value of the SG coupling,
where $q=-i$,
the $\slq$ symmetry corresponds to a topological extension of
$N=2$ supersymmetry;  away from this point the conserved currents
have fractional spin, and the $N=2$ symmetry algebra is deformed
into the $\slq$ algebra.  It is well known that the SG theory
possesses $sl(2)$ symmetry at a special value of the coupling
constant.  At this value of the coupling the $\slq$ symmetry
becomes the Yangian symmetry.  The Yangian is also a deformation
of affine Lie algebra, however the deformation preserves the
finite Lie subalgebra of the affine Lie algebra.  The conserved
currents generating the Yangian were studied by Bernard in
\ref\yang{\Yang}, based on some earlier work of L\"uscher\ref\rlus{\Luschi}.

What is remarkable about these symmetries is that they are large
enough to provide a non-perturbative solution to the S-matrix of
the theory.  Requiring that the conserved charges commute with the
S-matrix leads to a set of algebraic equations which characterize
the S-matrix up to an overall scalar factor.  This scalar can be fixed by
imposing crossing symmetry, unitarity, and the minimality criterion.
In the simple cases we are considering the S-matrices are
known\ref\Zamoi{\Zamolodchikovi}\ref\kath{\KaTh}.  These S-matrices
were originally found by imposing the Yang-Baxter equation, whereas
in our approach the solutions to the symmetry equations
automatically satisfy the Yang-Baxter equation.
This parallels what was accomplished by Drinfel'd and Jimbo in
their work on quantum groups\ref\drin{\Drinfeld}\ref\jimbo{\Jimbo} .
These authors understood that what underlies certain solutions of
the Yang-Baxter equation is a Hopf algebra.  Indeed the algebraic
symmetry equations we obtain are well-known equations in the
theory of quantum groups.

The framework based on non-local charges is applicable to
many more models than the ones considered in detail here.
The analysis of the SG theory extends readily to the
affine Toda theories and their Yangian limit\nlc .
More complicated examples include the many integrable
perturbations of rational conformal field theories.
For example a restriction of the quantum affine symmetry
of the SG theory can be used to derive the S-matrices
for the `$\Phi_{1,3}$' perturbations of the $c<1$ minimal
conformal models; this was developed in detail in \gio .
Quantum affine symmetry in perturbed minimal models was
also discussed by Mathur\ref\mathur{\Mat}.  The primary
distinction between the latter work and the results presented
here is that we deal only with genuine symmetries of a theory,
which means symmetries that follow from conserved currents
whose charges commute with the S-matrix.  The quantum group
structure in \mathur\ is not of this type, but rather is
more in the spirit of the quantum group symmetry of
conformal field theory, as presented in \ref\gos{\GoS}.

Since the non-local conserved charges virtually characterize
the S-matrix completely, it is natural to suppose that
they will provide non-perturbative off-shell information
as well.  Fields are completely characterized by their
form factors, i.e. matrix elements on the space of states
\ref\fed{\formn} .
For the usual internal symmetries based on finite dimensional
Lie algebras, the implication of the symmetry is that fields
are classified according to finite dimensional representations
of the Lie algebra.  In \form , this notion was generalized
to theories with an arbitrary Hopf algebra symmetry, by
introducing an adjoint action of the Hopf algebra on the
space of fields.  Similar algebraic constructions were made
in a lattice context in \ref\BF{D. Bernard and G.Felder,
`Quantum Group Symmetries in 2D Lattice Quantum Field Theory',
to appear in Nucl. Phys. B. }. For the case of infinite
Hopf algebra symmetry that we are interested in, the
fields were shown to comprise infinite dimensional
Verma-module representations.

I will begin by reviewing some general features of non-local
charges in 2d quantum field theory\yang\nlc .  In particular it will be
shown how the comultiplication for the charges arises from
the braiding of the currents with other fields.  This connection
was also observed by Gomez and Sierra in their study of
finite dimensional quantum group symmetry of conformal field
theory\gos .  I will then
construct explicitly the $\slq$ currents of the SG theory
and determine their algebra\nlc\gio .  The construction of \form\ will
then be presented.

\newsec{{\bf General Aspects of Non-Local Charges}}

Consider  some conserved currents $
\d_\mu J^a_\mu = 0$, defining some conserved charges
$Q^a = \inv{2\pi i} \int dx J^a_0 (x,t)$, and let us suppose
the following braiding relations with a set of fields $\Phi^i (x)$:
\eqn\IIxix{ J^a_\mu (x,t)\ \Phi^k(y,t) \ =\ \sum_{b,l}
 R^{ak}_{bl}\ \Phi^l(y,t)\ J^b_{\mu}(x,t) \qquad ;\qquad {\rm for} ~x<y . }
The action of the charges $Q^a$ on the fields, which we will denote
as $\ad {Q^a}$, can be defined as follows
\eqn\IIxx{ {\ad {Q^a} }\BL \Phi^k(y) \BR\ =\
\inv{2\pi i} \int_{\ga(y)}dz_{\nu}\ \ep^{\nu\mu}\ \jua(z)\ \Phi^k(y) ~,}
where the contour $\ga (y)$ begins and ends at $-\infty$ surrounding
the point $y$.
Using the
braiding relations \IIxix\ one derives
\eqn\IIxxi{ {\ad{Q^a}}\BL \Phi^k(y)\BR\ =\
Q^a\ \Phi^k(y) - R^{ak}_{bl}\ \Phi^l(y)\ Q^b .}
If $R^{ab}_{dc}$ is the braiding matrix of the currents with
themselves as in \IIxix , then the integrated version of \IIxxi\ is
\eqn\IIxxii{  \ad {Q^a} \( Q^b \) =
Q^a\ Q^b - R^{ab}_{dc}\ Q^c\ Q^d  . }
Under special conditions the contour on the right hand side of \IIxx\
can be closed to yield a new operator. In the case of
$\ad {Q^a} \( Q^b \)$ closure of the contour yields an algebra;  we will
encounter such a situation below.

The adjoint action of $Q^a$ on a product of two fields at
different spacial locations is again defined as in \IIxx ,
where now the contour surrounds the locations of both fields.
Using the braiding relations to pass the current through the
first field before acting on  the second, one finds that this
action has a non-trivial comultiplication
\eqn\IIxxiii{
\De \(  {Q^a} \) =  {Q^a} \ot 1 + \Theta^a_b \ot  {Q^b},}
where $\Theta^a_b$ is the braiding operator which acts on the
vector space spanned by the fields $\Phi^i$, i.e. $\Th^a_b$
has the matrix elements $\Th^{ak}_{bl} = R^{ak}_{bl}$.

\def\Php{\Phi_{{\rm pert.}}}
\def\php{\phi_{{\rm pert.}}}
\def\phb{{\bar \phi}_{{\rm pert.}}}
\def\zb{{\bar z}}

Let us apply the above  ideas to the non-local charges
that arise in a general perturbation of a conformal field theory
(CFT).
Consider a CFT  perturbed by a relevant operator
with zero Lorentz spin.
The perturbing field can be represented by
$\Php(z,\zb)= \php(z)\phb(\zb)$. The Euclidean action is taken to be
\eqn\action{S\ =\ S_{{\rm CFT}} + \frac{\la}{2\pi }
	\int d^2z\ \Php(z,\zb) }
where $\la$ is a generally dimensionful parameter that measures
the strength of the perturbation away from the conformal limit.
Here $z$ and $\zb$ denote  Euclidean
coordinates.
Chiral fields   $F(z,\zb)$, ${\bar F}(z,\zb)$
satisfy $\d_{\zb}\ F(z,\zb) = \d_z\ {\bar F}(z,\zb)=0$ in the conformal
limit.
Equations of motion for the perturbed chiral fields
which are local with respect to the perturbing field
can be deduced to first order
in perturbation theory  using Zamolodchikov's approach \ref\Zami{\Zamoiii}:
\eqn\motion{\eqalign{
\d_{\zb}\ F(z,\zb)\ &=\ \la \oint_z \frac{dw}{2\pi i}\
\Php(w,\zb)\ F(z) \cr
\d_z\ {\bar F}(z,\zb)\ &=\ \la \oint_{\zb} \frac{d{\bar w}}{2\pi i}\
\Php(z,{\bar w})\ {\bar F}(\zb) . \cr}}
Equations of motion to first order
can  be exact to all orders in perturbation theory, which can be seen
from scaling arguments.

Let us now suppose that there are
currents conserved to first order in perturbation theory:
\eqn\IIix{
\d_{\zb}\ J^a(z,\zb)\ =\ \d_z\ H^a(z,\zb),~~
\d_z\ {\bar J}^{\bar a}  (z,\zb)\ =\
\d_{\zb}\ {\bar H}^{\bar a} (z,\zb) . }
We assume that in the conformal theory these currents are chiral fields;
i.e. when $\la = 0$  they satisfy
$\d_{\zb}J^a=\d_z{\bar J}^{\bar a} =0$.
The condition for the currents to be conserved to first order in
perturbation theory is then a condition on the residue of the
operator product expansion (OPE) between them
and the perturbing field. Namely, the conservation laws \IIix{}\
hold if the residues of these
OPE's are total derivatives:
\eqna\IIx $$\eqalignno{
{\rm Res}_{z=w}\BL \php(w)\ J^a(z) \BR\ &=\ \d_z\ h^a(z) &\IIx a\cr
{\rm Res}_{\zb={\bar w}}\BL \phb({\bar w})\ {\bar J}^{\bar a}(\zb) \BR\
&=\ \d_{\bar z}\ {\bar h}^{\bar a}(\zb)  . &\IIx b \cr}$$
The conditions \IIx{}\ follow from Zamolodchikov's equation of motion \motion .
In  \IIix{}\ the fields $H^a$ are then
\eqn\eH{
H^a(z,\zb)= \la h^a(z)\phb(\zb) ,}
and similarly for ${\bar H}^{\bar a}(z,\zb)$.
{}From the conserved currents \IIix{}\ we define the conserved charges,
\eqn\eqc{
Q^a\ = \inv{2\pi i} \( \int dz J^a + \int d \zb H^a \) ,~~~
{\bar Q}^{\bar a}\ = \inv{2\pi i } \( \int d \zb {\bar J}^{\bar a}
+ \int dz {\bar H}^{\bar a} \) .  }

Since the currents $J^a$ and ${\bar J}^{\bar a}$ can be non-local,
we allow for non-trivial braiding  between them:
\eqn\IIxi{ J^a(x,t)\ {\bar J}^{\bar a}(y,t) \ =\
 R^{a{\bar a}}_{b{\bar b}}\ \ {\bar J}^{\bar b}(y,t)\ J^b(x,t)
\qquad ;\qquad x<y . }
There is no contradiction in having the above braiding relations
defined for all $x,y$;  we will give explicit examples in the sequel.
To find the commutation relations between
the charges $Q^a$ and ${\bar Q}^{\bar a}$
associated to these currents, we apply the general framework explained
above. Using  \IIx{}\ it is easy to compute
$\ad {Q^a} \( {\bar Q}^{\bar a} \)$ to lowest non-trivial order in perturbation
theory.
The result is:
\eqn\IIxiip{ Q^a\ {\bar Q}^{\bar a} - R^{a{\bar a}}_{b{\bar b}}\  \
{\bar Q}^{\bar b}\ Q^b\ =\
\frac{\la}{2\pi i }
 \int_t \BL dz\d_z + d\zb \d_{\zb}\BR \  h^a(z){\bar h}^{\bar a}(\zb) . }
The right hand side of the above equation is a topological charge,
and may be understood as a generalization of the topological
extensions of  supersymmetry in two-dimensions
\ref\WO{\WOlive}.

\newsec{{\bf The $\slq$ Currents in the Sine-Gordon Theory }}

\medskip
\def\bh{\hat{\beta}}
\def\hb{\bh}

I  now review how the \slqh loop algebra symmetry is realized
in the sine-Gordon theory\nlc .  The SG theory may be treated
as a massive perturbation of the $c=1$ conformal field theory
corresponding to a single real scalar field:
\eqn\IIxii{ S\ =\ \inv{4\pi}\int d^2z\ \d_z\Phi \d_{\zb}\Phi \
+ \frac{\la}{\pi }\ \int d^2z\ :\cos\(\hb \Phi\): ~~.}
The theory has a well-known conserved topological current
$$\CJ_\mu^{\rm top.} (x) = \frac{\bh}{2\pi} \ep_{\mu\nu} \d_\nu \Phi (x) $$
which defines the topological charge
\eqn\IIxiii{
T= \frac{\bh}{2\pi} \int_{-\infty}^{\infty} dx \> \d_x \Phi (x). }
With
the above $\bh$-dependent normalization of the topological charge,
the soliton states are known to have $T=\pm 1$ \ref\rcole{\Colemani}
\ref\rmand{\Mand}.

Define the quasi-chiral components $\vphi , \vphib$ of $\Phi$ as
\eqn\IIxiv{\eqalign{
\vphi(x,t)\ &=\ \half\BL\Phi(x,t)
+\int_{-\infty}^x dy\ \d_t\Phi(y,t)\BR \cr
\vphib(x,t)\ &=\ \half\BL\Phi(x,t)
-\int_{-\infty}^x dy\ \d_t\Phi(y,t)\BR ,\cr}}
such that $\Phi = \vphi + \vphib$.  When $\lambda =0$,
$\vphi = \vphi (z)$, and
$< \vphi (z) \vphi (w) > = -\log (z-w).$
Similarly for $\vphib$.
We will make use of the following braiding relations:
\eqn\IIxivb{\eqalign{
\exp\(ia\vphi(x,t)\)\ \exp\(ib\vphi(y,t)\)\ &=\
e^{\pm i\pi ab}\ \exp\(ib\vphi(y,t)\)\ \exp\(ia\vphi(x,t)\)
\ ;\ {\rm for}\ x{> \atop <}y \cr
\exp\(ia\vphib(x,t)\)\ \exp\(ib\vphib(y,t)\)\ &=\
e^{\mp i\pi ab}\ \exp\(ib\vphib(y,t)\)\ \exp\(ia\vphib(x,t)\)
\ ;\ {\rm for}\ x{> \atop <}y \cr
\exp\(ia\vphi(x,t)\)\ \exp\(ib\vphib(y,t)\)\ &=\
e^{ i\pi ab}\ \exp\(ib\vphib(y,t)\)\ \exp\(ia\vphi(x,t)\)
\ ;\ \forall\ x, y .  \cr
}}
The topological charge of these fields follows from the relation
\eqn\IIxivc{
\[ T, \exp \( ia\vphi + ib\vphib \) \] = \bh (a-b)
\exp \( ia \vphi + ib\vphib \) .}
The conformal dimensions $(h,\bar{h} )$ of the field
$\exp \( ia\vphi + ib\vphib \)$ are
$(a^2 /2 , b^2 /2 )$, and its Lorentz spin (eigenvalue under
Lorentz boosts) is $h-\bar{h}$.

Using the general results of the last section,
it can be shown that the model \IIxii\ has the following non-local
quantum conserved currents:
$\d_{\zb}J_{\pm}\ =\ \d_zH_{\pm}
 ; ~~
\d_z{\bar J}_{\pm}\ =\ \d_{\zb}{\bar H}_{\pm} ,$
where
\eqna\IIxvi
$$\eqalignno{
J_{\pm}(x,t)&=\exp\(\pm\frac{2i}{\hb}\ \vphi(x,t)\), ~~~~~
{\bar J}_{\pm}(x,t)=\exp\(\mp\frac{2i}{\hb}\ \vphib(x,t)\)
	 &\IIxvi a\cr
H_{\pm}(x,t)&=\la
\frac{\bh^2}{\bh^2 -2}
\exp\lb \pm i\(\frac{2}{\hb}-\hb\)\vphi(x,t)
	\mp i\hb\ \vphib(x,t)\rb
&\IIxvi b\cr
{\bar H}_{\pm}(x,t)&=\la
\frac{\bh^2}{\bh^2 -2}
\exp\lb \mp i\(\frac{2}{\hb}-\hb\)\vphib(x,t)
	\pm i\hb\ \vphi(x,t)\rb .
&\IIxvi c\cr
  }$$
The coefficient $2/\bh$ in the exponent for the
 components $J_\pm , \Jb_\pm$ of the currents follows
 from the residue condition \IIx{}.

These currents define the conserved  charges $Q_\pm , \Qb_\pm$.
The conserved charges have non-trivial Lorentz spin:
\eqn\IIxviii{\inv{\ga}\ \equiv\ {\rm spin}\(Q_{\pm}\)\ =\
-{\rm spin}\({\bar Q}_{\pm}\)\ =\
\frac{2}{\hb^2}-1 .}
Consequently the non-local conserved currents have non-trivial
braiding relations among themselves and with other fields.
In particular the braiding relations
\IIxivb{} \ imply:
\eqn\IIIxxip{\eqalign{
J_{\pm}(x,t)\ {\bar J}_{\mp}(y,t)\ &=\
q^{-2}\ {\bar J}_{\mp}(y,t)\ J_{\pm}(x,t) \qquad ;\ \forall\ x,y \cr
J_{\pm}(x,t)\ {\bar J}_{\pm}(y,t)\ &=\
q^{2}\ {\bar J}_{\pm}(y,t)\ J_{\pm}(x,t)\qquad;\ \forall\ x,y \cr}}
where
\eqn\eqd{ q=\exp(-2\pi i /\hb^2)=-\exp(-i\pi/\ga).}

\def\bQ{ {\bar Q} }

Using the above braiding relations and the  equation \IIxiip\
one finds that the conserved charges satisfy the
relations
\eqnn\IIIxxii  $$\eqalignno{
Q_{\pm}\ \bQ_{\pm} - q^{2}\ \bQ_{\pm}\ Q_{\pm}\ &=\ 0 &\IIIxxii \cr
Q_{\pm}\ \bQ_{\mp} - q^{-2}\ \bQ_{\mp}\ Q_{\pm}\ &=\
\frac{\la}{2\pi i}\ \ga^2 \int_t dx\ \d_x
\lb\exp\(\pm i\(\frac{2}{\hb}-\hb\)\Phi(x,t)\)\rb .
\cr}$$
The topological charges on the right hand side of \IIIxxii{}\
can be expressed in terms of the usual topological charge $T$
in \IIxiii . A soliton configuration can be taken to satisfy
$\Phi(x=\infty)=0$; the classical soliton solutions do in fact satisfy
this. Integrating \IIIxxii{}\ we obtain the algebra to lowest
non-trivial order in perturbation theory:
\eqna\IIIxxiii  $$\eqalignno{
Q_\pm\ \bQ_\pm  - q^{2}\ \bQ_\pm\ Q_\pm  \ &=\ 0 &\IIIxxiii a\cr
Q_\pm \ \bQ_\mp  - q^{-2}\ \bQ_\mp \ Q_\pm  \ &=\
a \BL 1- q^{\pm 2T}\BR
&\IIIxxiii b\cr
\BBL\ T\ ,\ Q_{\pm}\ \BBR\ =\ \pm2\ Q_{\pm} ,~~&~~
\BBL\ T\ ,\ \bQ_{\pm}\ \BBR\ =\ \pm2\ \bQ_{\pm}
&\IIIxxiii c\cr
\cr}$$
where $a\equiv \la \ga^2 /2\pi i $.
In deriving the last equations we have used  \IIxivc .
It can be shown using scaling arguments that this
algebra is exact to all orders.
Note that when $q=-i$, ($\hb=2/\sqrt{3}$), the algebra
\IIIxxiii{}\ is a topological extension of the $N=2$ supersymmetry algebra.

The above algebra \IIIxxiii{}\ is isomorphic to the $\slq$ loop algebra.
Let $E_i,\ F_i,\ H_i$, $i=0,\ 1$, denote the Chevaley basis for the
$\slq$ algebra.
They satisfy the following
defining relations:
\eqn\rel{\eqalign{
\BBL\ H_i\ ,\ E_j\ \BBR\ &=\ a_{ij}\ E_j\cr
\BBL\ H_i\ ,\ F_j\ \BBR\ &=\ -a_{ij}\ F_j\cr
\BBL\ E_i\ ,\ F_j\ \BBR\ &=\ \de_{ij}\ \frac{q^{H_i}-q^{-H_i}}{q-q^{-1}}\cr}}
with $a_{ij}$ the Cartan matrix of the affine Kac-Moody algebra
${\hat {sl(2)}}$\foot{$a_{11} = a_{00} = -a_{10}= -a_{01} = 2$}.
The relations between the non-local charges $Q_{\pm}$ and $\bQ_{\pm}$
and these generators are:
\eqn\IIIxxiv{\eqalign{
Q_+\ =\ c\ E_1\ q^{H_1/2} \qquad &;\qquad
Q_-\ =\ c\ E_0\ q^{H_0/2} \cr
\bQ_-\ =\ c\ F_1\ q^{H_1/2} \qquad &;\qquad
\bQ_+\ =\ c\ F_0\ q^{H_0/2} \cr
 T\ =\ H_1\ &=\ -H_0 \cr}}
where $c$ is a constant ($c^2=\frac{\la}{2\pi i} \gamma^2 (q^{-2}-1)$).
The last equation in \IIIxxiv\ reflects the fact that the center of
$\slq$ is zero, and we are actually dealing with a deformation of
the loop algebra.

\def\a{{\rm ad}}
The complete set of relations for the $\slq$ algebra include the
additional deformed Serre relations.
They are in this case
\eqn\serviii{
{\rm ad}_{Q_\pm}^{\dis ~3} \( Q_\mp \) = {\rm ad}_{\Qb_\pm}^{\dis ~3}
\( \Qb_\mp \)
= 0 . }
These relations were proven in \gio\ by studying the conditions
underwhich the contours in the
 expression
$\a_{Q_+}^{\dis ~n} \( J^-_\mu (y,t) \) $ could be closed.
Using the braiding relations of the currents one finds they
can be closed precisely when $n=3$.  Using scaling arguments
and the short distance properties of the ultraviolet CFT,
one then deduces that the operator obtained upon closing the
contours is necessarily zero.

I will now describe how one can use the above $\slq$ symmetry to
derive the S-matrix for the scattering of solitons.  To do this
one needs to know how the symmetry algebra is represented on
the space of asymptotic states.
The asymptotic soliton states of topological charge $\pm 1$
are denoted $\ket{\pm 1/2,\th}$, where $\th$ is the rapidity:
\eqn\IIxxvii{
p_0 (\th ) = m \cosh (\th ) ~~~~~~~~p_1 (\th ) = m \sinh (\th ) , }
and $T \ket {\pm 1/2 , \th } = \pm \ket{\pm 1/2 , \th } $.
A set of chiral fields of topological charge $\pm 1$ with
non-vanishing matrix elements between the states and the vacuum
can be taken to be
\eqn\IIxxviib{
 \Psi_{\pm}(x,t)\ =\ \exp\(\pm\frac{i}{\hb}\ \vphi(x,t)\), ~~~~
 \Psib_{\pm}(x,t)\ =\ \exp\(\mp\frac{i}{\hb}\ \vphib(x,t)\) .  }
The fields $\Psi_\pm$ and $\Psib_\pm$ do not create independent
particle states, for the usual reasons.
The representation of \slqh on the space of one-particle states
can be shown to be
\eqn\IIxxviii{\eqalign{
Q_{\pm}\ &=\ c\ e^{\th/\ga}\ \sigma_{\pm}\ q^{\pm \sigma_3 /2} \cr
\bQ_{\pm}\ &=\ c\ e^{-\th/\ga}\ \sigma_{\pm}\ q^{\mp \sigma_3 /2} \cr
T &= \sigma_3 ,  \cr}}
where $\sigma_\pm , \sigma_3$ are the Pauli spin matrices.
The on-shell operators $\exp (\th /\ga )$ are a consequence of the
Lorentz spin of the conserved charges, since the Lorentz boost
generator is represented as $-\d_\th $ on-shell.
The above representation of the \slqh loop algebra is in the so-called
principal gradation.

\def\S{{\hat{S}}}

The action of the conserved charges on the multiparticle states is
provided by the comultiplication which follows from \IIxxiii\ and
the braiding of the non-local currents with the soliton fields.
{}From the braiding relations
\eqn\IIIxxxviii{\eqalign{
J_{\pm}(x,t)\ \Psib_{T}(y,t)\ &=\ q^{\pm T}\
\Psib_{T}(y,t)\ J_{\pm}(x,t)\quad;\quad \forall\ x,y \cr
{\bar J}_{\pm}(x,t)\ \Psi_{T}(y,t)\ &=\ q^{\mp T}\
\Psi_{T}(y,t)\ {\bar J}_{\pm}(x,t)\quad;\quad \forall\ x,y ~~,\cr}}
one thereby deduces that the comultiplication is:
\eqna\IIIxxxix  $$\eqalignno{
\De\(Q_{\pm}\)\ &=\ Q_{\pm}\otimes 1 + q^{\pm T}\otimes Q_{\pm}
&\IIIxxxix a\cr
\De\(\bQ_{\pm}\)\ &=\ \bQ_{\pm}\otimes 1 + q^{\mp T}\otimes \bQ_{\pm}
&\IIIxxxix b\cr
\De\(T\)\ &=\ T\otimes 1 + T\otimes 1.
&\IIIxxxix c\cr}$$
and is equivalent to the usual one. The two-particle to two-particle
S-matrix $\S$ is an operator from $V_1 \ot V_2$ to $V_2 \ot V_1$,
where $V_i$ are the vector spaces spanned by $\ket{ \pm 1/2 , \th_i }$.
The \slqh symmetry of the S-matrix is the condition
\eqn\IIxxixa{
\S_{12}
(\th_1 - \th_2 ;q ) \,\De_{12} (a) = \De_{21}
(a) \, \S_{12}  (\th_1 -\th_2 ;q) ~~~~~~a\in
\CU_q \( \hat{sl(2)} \) .}
These equations are familiar in the theory of quantum groups, and
show that the S-matrix is just the universal $\CR$-matrix
specialized to the rapidity-dependent representations of
$\slq$.
The minimal solution to these symmetry equations is the conjectured
S-matrix of SG solitons\Zamoi .

\def\a{\CA}

\newsec{{\bf General Hopf Algebra Symmetry of Fields}}

We have seen how the infinite dimensional quantum groups are
represented on the space of asymptotic states though finite
dimensional rapidity-dependent representations.  In this section
I will address how the symmetry is realized on the space of
fields of the theory.  We already described how to act on
fields with conserved charges in \IIxx .  In order to determine
how the fields obtained via this adjoint action transform under the
symmetry, we first express this adjoint action in a more
suitable way\form .
Let $\CA$ be a Hopf algebra equipped with
comultiplication $\Delta : \CA \to \CA \ot \a$,
counit $\ep : \a \to \C$, and antipode $s: \a \to \a$,
with the following properties:
\eqna\IIiii
$$\eqalignno{
\De (a) \De (b) &= \De (ab) &\IIiii{a}\cr
\( \De \ot id \) \De (a) &= \( id \ot \De \) \De (a) &\IIiii{b} \cr
\( \ep \ot id \) \De (a) &= \( id \ot \ep \) \De (a) = a &\IIiii{c} \cr
m \( s\ot id \) \De (a) &= m \( id\ot s \) \De (a) = \ep (a) &\IIiii{d} \cr
}$$
for $a,b\in \a$ and $m$ the multiplication map: $m(a\ot b) = ab$.
Eq. \IIiii{a}\ implies $\De$ is a homomorphism of $\a$ to $\a \ot \a$,
\IIiii{b}\ is the coassociativity, and \IIiii{c,d}\ are the defining
properties of the counit and antipode.  Let us chose a basis
$\{ e_a \}$ in $\CA$, and define the multiplication and
comultiplication in terms of structure constants $m_{ab}^c$ and
$\mu^{ab}_c$:
\eqna\Iii
$$\eqalignno{
e_a \>  e_b &= \m abc \; e_c &\Iii {a} \cr
\De (e_a ) &= \u bca \; e_b \ot e_c &\Iii {b} \cr }$$

\def\a{\CA}

For our purposes we will also need the concept of
the quantum double of $\CA$, which is a construction
due originally to Drinfel'd
\ref\rdrinb{V. G. Drinfel'd, `Quantum Groups',
{\it Proceedings of the International Congress of
Mathematicians}, Berkeley, CA, 1986.}
\ref\rfadd{L. Faddeev, N.  Reshetikhin, and L. Takhtajan,
in {\it Braid Group, Knot Theory and Statistical Mechanics},
C . N. Yang and M. L. Ge (ed.),  World Scientific, 1989. }.
Let $\a^*$ denote a Hopf
algebra dual to $\CA$ in the quantum double sense, with
basis $\{ e^a \} $.  The elements of the dual basis are
defined to satisfy the following relations:
\eqna\Iiv
$$\eqalignno{
e^a e^b &= \u abc \> e^c &\Iiv {a} \cr
\De (e^a ) &= \m bca \> e^c \ot e^b &\Iiv {b} .\cr}$$
The quantum double is a Hopf algebra structure on the space
$\a \ot \a^* $. The relations between the elements
$e_a$ and the dual elements $e^a $ are defined as follows:
\eqna\Ix
$$\eqalignno{
e_a \, e^b &= \mu^{kcl}_a \> m^b_{idk} \> {s'}^i_l ~e^d \, e_c &\Ix {a} \cr
e^b \, e_a &= \mu_a^{lck} \> m_{kdi}^b \> {s'}^i_l ~e_c \, e^d  &\Ix {b} \cr
}$$
where
$\mu^{lck}_a = \u lia \> \u cki ; ~~
m^b_{kdm} = \m kdi \> \m imb $, and $s'(e_a ) = s'^b_a e_b$ is the
skew antipode.
The  universal
$\CR$-matrix is an element of $\a \ot \a^*$ defined by
\eqn\Ivii{
\CR = \sum_a e_a \ot e^a , }
satisfying
\eqn\Iviii{
\CR ~ \De (e_a ) = \De' (e_a ) \CR ,}
for all $e_a$, where $\De'$ is the permuted comultiplication.
$\CR$ also satisfies the Yang-Baxter equation.

Let us now define the adjoint action on a field or product of
fields as
\eqn\IIIxiv{
\ad {e_a} \bigr( \Phi (x_1 ) \cdots \Phi (x_n ) \bigl) =
\mu_a^{bc}
e_b \ \Phi (x_1 ) \cdots \Phi (x_n ) \ s(e_c) . }
This adjoint action generalizes the ordinary commutator in
Lie algebra symmetry
to an arbitrary Hopf algebra, and has the appropriate
properties.
In particular\foot{
The adjoint action in \IIIxiv\ differs
slightly from the one used in \form .  The origin of the
difference is the same as `right' versus `left' action
in Lie group theory, i.e. the adjoint action in \form\ satisfied
$\ad {e_a} \ad {e_b} = \ad {e_b e_a}$.},
\eqn\adprop{
\ad {e_a}  \ad {e_b}  = \ad {e_a e_b}  . }
This latter property implies that
fields related through adjoint
action form a representation of $\CA$.  More precisely let
$\Phi_\Lambda (x)$ denote the set of fields so obtained,
and let $\rho_\La (a)$ denote the  representation of
$\CA$.  Then
\eqn\IIIxv{
\ad {e_a}   \( \Phi_\La (x) \) = \rho_\La (e_a ) \Phi_\La (x). }
The adjoint action also satisfies the important property
\eqn\IIIxvi{
\ad {e_a} \bigl( \Phi_{\La_1} (x_1 ) \> \Phi_{\La_2} (x_2 ) \bigr)
= \mu_a^{bc} \ad {e_b} \bigl( \Phi_{\La_1} (x_1 ) \bigr)
\ad {e_c}  \bigl( \Phi_{\La_2} (x_2 ) \bigr) . }

\noindent
These properties are all proven using  the Hopf algebra properties \IIiii{}.

An important consequence of \IIIxvi\ is that the
braiding of the multiplets of fields is given by the universal
$\CR$-matrix:
\eqn\IIxxiz{
\Phi_{\hw_2} (y,t) \> \Phi_{\hw_1} (x,t)
= \CR_{\rhwo 1 ,\rhwo 2 } \> \Phi_{\hw_1 } (x) \> \Phi_{\hw_2 } (y)
{}~~~~~x<y, }
where $\CR_{\rhwo 1 ,\rhwo 2 } $ is the universal $\CR$ matrix
specialized to the representations $\rhwo{{1,2}}$ of the fields
$\Phi_{\hw_{1,2}}$.  The above relation is easily proven by
applying $\ad {e_a}$ to both sides of \IIxxiz , and using \IIIxvi\
to prove that $\CR$ must satisfy its defining relations \Iviii .

It can be shown that the above abstract adjoint action is
equivalent to the adjoint action defined via the contours in
the quantum field theory \IIxx .
The proof goes as follows\form .  We first define an adjoint
representation $\rho_{adj}$ of $\a , \a^*$:
\eqna\IIxxiii
$$\eqalignno{
\bra b \rho_{{\rm adj}} (e_a ) \ket c &= \m acb  &\IIxxiii {a} \cr
\bra c \rho_{{\rm adj}} (e^a) \ket b &= \u abc &\IIxxiii {b} .\cr
}$$
That the structure constants form a representation of the algebra
is a consequence of the associativity and coassociativity of
$\CA$.  We associate the conserved currents $J^\mu_a (x)$ for
the charge $e_a$
to the fields in $\Phi_{\rho_{adj}} (x)$.
 The braiding of
these currents with other fields can then be deduced from \IIxxiz  .
One finds
\eqn\IIxxiv{
J_a^\mu (y) ~ \Phi_\hw (x) = \rhw \( R^b_a \) \Phi_\hw (x) ~ J^\mu_b (y)
{}~~~~~x<y , }
where
\eqn\IIxxv{
R^b_a \equiv \u cba \> e_c . }
Using these braiding relations in \IIxxi , one finds that
\IIxxi is equivalent to \IIIxiv .

For the infinite dimensional Hopf algebras we are considering, the
multiplets of fields $\Phi_\La (x)$ can be seen to
comprise infinite dimensional Verma-module representations.
This was shown in \form\ by studying the form-factors of
these fields.  Consider the matrix element of a field
$\phi (x)$ on the space of asymptotic states (form factor):
\eqn\IIix{
\bra {\al_1 , ..., \al_m } \> \phi (0) \> \ket {\be_n ,..., \be_1}
,   }
where $\al_i, \be_i$ are rapidites.
Given the representation of $\CA$ on the asymptotic states, it is
clear that the form factor of any other field obtained from adjoint
action on $\phi$ is explicitly computable from the knowledge of
the form factors of $\phi (x)$, since the elements of
$\CA$ on the left or right of the field in \IIIxiv\ give a known
transformation on states.

Let me illustrate this for the case of the Yangian symmetry.
The SG theory at the coupling $\bh = \sqrt{2}$ is equivalent to
the chiral sl(2) Gross-Neveu model, and the spectrum consists of
a doublet of solitons.  At this point the S-matrix
is sl(2) invariant, and is a well-known rational solution of
the Yang-Baxter equation:
\eqn\IIIii{
S_{12} (\theta ) = s_0 (\theta ) \( \be  - i\pi P_{12} \), }
where $P_{12}$ is the permutation operator:
\eqn\IIIiii{
P_{12} = \inv{2} \( \sum_{a=1}^3 \sigma^a \ot \sigma^a + 1 \) , }
($\sigma^a $ are the Pauli spin matrices), and
\eqn\IIIiv{
s_0 (\be ) = \frac{ \Gamma (1/2 + \be /2\pi i ) \> \Gamma (-\be /2\pi i ) }
{\Gamma (1/2 - \be /2\pi i ) \> \Gamma (\be /2\pi i ) }  . }

Let $Q_0^a, a= 1,2,3$ denote the global sl(2)
generator.  The S-matrix is invariant under some additional symmetries
generated by $Q_1^a$, which are represented on the asymptotic 1-particle
states as $Q_1^a |\theta > = \theta t^a |\theta >$, where
$t^a$ is a 2-dimensional representation of sl(2).
These charges satisfy the algebra
\eqn\alg{ \[ Q_0^a , Q_0^b \] = f^{abc} Q_0^c , ~~~~~
\[ Q_0^a , Q_1^b \] = f^{abc} Q_1^c . }
They have the following comultiplication:
\eqna\IIIviii
$$\eqalignno{
\De (\qo a ) &= \qo a \ot 1 + 1\ot \qo a &\IIIviii {a} \cr
\De (\qp a ) &= \qp a \ot 1 + 1\ot \qp a + \al \> \f abc \> \qo b \ot \qo c ,
&\IIIviii {b} \cr }$$
where
$\al = -i\pi$.  The extra term in the comultiplication
for $Q_1^a$ is required for invariance of the S-matrix.

These are the defining relations of the Yangian.
This symmetry  may be thought of as arising in the $q \to -1$ limit of
the $\slq$ symmetry.
The Yangian, as originally defined by Drinfel'd\drin , is
a deformation of half of the $\hat{sl(2)}$ affine Lie algebra
that preserves the finite sl(2) subalgebra.  In order to
determine the universal $\CR$-matrix, it was necessary in
\form\ to extend the Yangian to its quantum double.
It was then shown  that the
deformation of the remaining half of the affine Lie algebra  resides
in this quantum double.

By studying the form factors of the adjoint action of
$Q_1^a$ on the energy-momentum tensor $T_{\mu\nu}$ one
can derive the fundamental identity:
\eqn\IIIxxxix{
\[ \qp a , T_{\mu\nu} (x) \] = -\inv{2}
\( \ep_{\mu\al } \d_{\al} J^a_\nu (x)
+ \ep_{\nu\al} \d_\al J^a_\mu (x) \) , }
where $J^a_\mu (x)$ is the global current for the
sl(2) charge $Q_0^a$.  Thus we begin to see how non-trivial
the Yangian symmetry is:  the energy-momentum tensor and
global current share the same Yangian multiplet.

\newsec{{\bf Conclusions}}

It is clear from the above discussion that the infinite
quantum group symmetry of integrable massive models
can provide new non-perturbative results.  I believe it
is possible to completely define the models and determine
their main properties from these symmetries alone.
However much remains to be understood toward the completion
of this program.
In particular the form factors have not been completely understood
purely in terms of the symmetry, unlike the S-matrix.
Some steps in this direction have been taken in the
works\ref\kol{I. Frenkel and N. Yu. Reshetikhin, in
preparation, talk given at this conference.}
\ref\kz{F. A. Smirnov, `Dynamical Symmetries of Massive
Integrable Models', RIMS preprint.}, where it was shown
how one of the axioms of the form factor bootstrap
can be understood as a deformed Knizhnik-Zamolodchikov
equation.  The issue of how the complete set of fields
in a theory is classified according to the symmetry is
another important problem that has not been completely
solved, and this information is necessary for a formulation
of the correlation functions.

\bigskip

\centerline{Acknowlegements}

I am especially grateful toward my collaborators on this work,
D. Bernard, G. Felder, and F. Smirnov.  I also thank the organizers
of this conference for the opportunity to present these results.

%\figures
%\fig{1}{bla}
\listrefs
%\end
\bye